\documentclass[11pt,twoside]{article}
\usepackage{asp2006}
\usepackage{epsf}
\usepackage{psfig}
\usepackage{graphicx}
\usepackage{lscape}

\begin{document}
\title{Keeping the Universe ionised: photoheating and the high-redshift
 clumping factor of the intergalactic gas} 
\author{Andreas H. Pawlik, \altaffilmark{1,}\altaffilmark{2} Joop Schaye,
  \altaffilmark{2} and Eveline van Scherpenzeel\altaffilmark{2}}  
\altaffiltext{1}{Department of Astronomy, University of Texas, Austin, TX 78712, USA}   
\altaffiltext{2}{Leiden Observatory, Leiden University, P.O. Box 9513, 2300 RA Leiden, The Netherlands} 
\begin{abstract} 
The critical star formation rate density required to keep the intergalactic
hydrogen ionised depends crucially on the average rate of recombinations in
the intergalactic medium (IGM). This rate is proportional to the clumping
factor $C_{\rm IGM} \equiv \langle \rho_{\rm{b}}^2 \rangle_{\rm IGM} /
\bar\rho_{\rm{b}}^2$, where $\rho_{\rm{b}}$ and $\bar\rho_{\rm{b}}$ are the
local and cosmic mean baryon density, respectively and the brackets $\langle
\rangle_{\rm IGM}$ indicate spatial averaging over the recombining gas in the
IGM.  We perform a suite of cosmological smoothed particle hydrodynamics
simulations to calculate the volume-weighted clumping factor of the IGM at
redshifts $z \ge 6$. We investigate the effect of photoionisation heating by a
uniform ultraviolet background and find that photoheating strongly reduces the
clumping factor as the increased pressure support smoothes out small-scale
density fluctuations. Even our most conservative estimate for the clumping
factor, $C_{\rm IGM} = 6$, is five times smaller than the clumping factor that
is usually employed to determine the capacity of star-forming galaxies to keep
the $z \approx 6$ IGM ionised. Our results imply that the observed population
of star-forming galaxies at $z \approx 6$ may be sufficient to keep the IGM
ionised, provided that the IGM was reheated at $z \ga 9$ and that the fraction
of ionising photons that escape star-forming regions to ionise the IGM is
larger than $0.25$.
\end{abstract}
\section{Introduction}
The absence of a Gunn-Peterson trough in many of the absorption spectra
towards high-redshift quasars suggests that the reionisation of intergalactic
hydrogen was completed at redshifts $z \ga 6$
(e.g., \citealp{pawlik_Fan:2006}).  Current observational estimates of the
ultraviolet (UV) luminosity density at redshifts $z \la 6$ (for a
comprehensive discussion see, e.g., \citealp{pawlik_Bouwens:2007}), 
on the other
hand, may imply star formation rate (SFR) densities several times lower than
the critical SFR density required to keep the intergalactic medium (IGM)
ionised.
\par
The critical SFR density,
\begin{equation}
  \dot{\rho}_* \approx  0.027 ~\mbox{M}_{\odot} ~\mbox{yr}^{-1} ~\mbox{Mpc}^{-3} {f_{\rm{esc}}}^{-1} \left (
  \frac{C_{\rm IGM}}{30} \right )  \left ( \frac{1+z}{7} \right )^3  \left ( \frac{\Omega_{\rm{b}}h_{70}^2}{0.0465} \right )^2,
  \label{Eq:CriticalSfr}
\end{equation}
originally derived by \cite{pawlik_Madau:1999} and here rescaled to match the
most recent WMAP estimate for the cosmic baryon density
(\citealp{pawlik_Komatsu:2008}), is inversely proportional to the escape fraction
$f_{\rm{esc}}$, i.e.  the fraction of ionising photons produced by
star-forming galaxies that escape the interstellar medium (ISM) to ionise the
IGM. It is proportional to the average recombination rate in the IGM,
parametrised using the dimensionless clumping factor $C_{\rm IGM}\equiv
\langle \rho_{\rm{b}}^2\rangle_{\rm IGM} / \bar\rho_{\rm{b}}^2$, where
$\rho_{\rm{b}}$ is the baryon density, $\bar\rho_{\rm{b}}$ is the mean baryon
density of the Universe and the brackets $\langle \rangle_{\rm IGM}$ indicate
spatial averaging over the gas constituting the recombining IGM.
\par
Most observational studies that compare the SFR density derived from estimates
of the UV luminosity density at redshift $z \approx 6$ to the critical SFR
density assume an escape fraction $f_{\rm esc} \la 0.5$ and a clumping factor
$C_{\rm IGM} = 30$.  Consequently, the observed population of galaxies has
been found to be incapable of keeping the intergalactic hydrogen ionised,
forming massive stars at a rate which is up to an order of magnitude lower
than required by Eq.~\ref{Eq:CriticalSfr}. It has, however, been noted that
the discrepancy between the observationally inferred and critical SFR
densities at $z \approx 6$ could be resolved if the employed clumping factor
were too high (e.g., \citealp{pawlik_Sawicki:2006}).
\par
In this work we perform cosmological Smoothed Particle Hydrodynamics (SPH)
simulations to compute the clumping factor of the IGM. We pay particular
attention to the fact that the clumping factor depends on the definition of
which gas comprises the IGM. By comparing the critical SFR density, updated
using our estimate of the clumping factor, with the observationally inferred
SFR density we argue that the observed population of UV galaxies may well be
capable of keeping the $z \approx 6$ Universe ionised. We also investigate the
effect of photoionisation heating by a uniform UV background on the evolution
of the clumping factor.~We demonstrate that photoheating significantly lowers
the clumping factor.~Our results are insensitive to the redshift at which the
UV background is turned on.
\par  
This work is described in more detail in \cite{pawlik_Pawlik:2009a}. 
\section{Simulations}
\label{Sec:Clumping:Simulations}
We use a modified version of the N-body/TreePM/SPH code {\sc gadget-2}
(\citealp{pawlik_Springel:2005}) to perform a suite of cosmological SPH
simulations including radiative cooling.  We assume a flat $\Lambda$CDM
universe and employ the set of cosmological parameters $[\Omega_{\rm{m}},
\Omega_{\rm{b}}, \Omega_\Lambda, \sigma_8, n_{\rm{s}}, h]$ given by $[0.258$,
$0.0441$, $0.742$, $0.796$, $0.963$, $0.719]$, consistent with the {\it WMAP}
5-year results (\citealp{pawlik_Komatsu:2008}).
\par
The gas is of primordial composition, with a hydrogen mass fraction $X =
0.752$ and a helium mass fraction $Y = 1-X$.  Radiative cooling and heating
are included assuming ionisation equilibrium, as described in
\cite{pawlik_Wiersma:2009}. Molecular hydrogen and deuterium and their
catalysts are kept photo-dissociated by a soft UV background and never
contribute to the cooling rate (e.g., \citealp{pawlik_Haiman:1997}).  Star
formation is modelled by turning gas particles into star particles using the
recipe of \cite{pawlik_Schaye:2008}.
\par
All simulations use $256^3$ DM particles and $256^3$ gas particles in a box of
comoving size $L = 6.25 ~h^{-1}~\mbox{Mpc}$.  We perform simulations including
photoionisation by a uniform, evolving \cite{pawlik_Haardt:2001} UV
background in the optically thin limit at redshifts below the {\it reheating}
redshift\footnote{If $z_{\rm{r}} > 9$, we use the $z = 9$
\cite{pawlik_Haardt:2001} UV background for all redshifts $ 9 < z \le
z_{\rm{r}}$, and employ the evolving \cite{pawlik_Haardt:2001} UV background
for redshifts $z \le 9$.} $z_{\rm{r}}$. These simulations are denoted
\textit{r[$z_{\rm{r}}$]L6N256}. To study the effect of photoheating, we
compare these simulations to a simulation that does not include
photoionisation and which is denoted \textit{L6N256}. We explore a range of
reheating redshifts to investigate the sensitivity of our conclusions to
changes in this parameter. We note that for the simulations that included the
UV background the employed box size is sufficiently large and the employed
resolution is sufficient high to obtain converged results 
(for details see \citealp{pawlik_Pawlik:2009a}).
\par
\section{The clumping factor}
\label{Sec:ClumpingFactor}
\par
Our main motivation for computing the clumping factor of the IGM is to
evaluate the critical SFR density required to keep the IGM ionised.  The
critical SFR density describes a balance between the number of ionising
photons escaping into the IGM (parametrised by $f_{\rm esc}$) and the number
of ionising photons that are removed from the IGM due to photoionisations of
recombining hydrogen ions (parametrised by $C_{\rm IGM}$). It is important to
realize that only recombinations leading to the removal of ionising photons
which escaped the ISM of the star-forming regions contribute to this balance.
\par
We employ a threshold density criterion to separate the gas in the ISM from
the gas in the IGM (e.g., \citealp{pawlik_Miralda:2000}). Ionising photons are
counted as escaped once they enter regions with gas densities $\rho_{\rm b} <
\rho_{\rm thr}$. We treat the threshold density $\rho_{\rm thr} \equiv
\bar\rho_{\rm b} \Delta_{\rm thr}$ as a parameter and compute the clumping
factor $C(< \Delta_{\rm thr}) \equiv \langle \rho_{\rm{b}}^2 \rangle_{\rm
\rho_{\rm b} < \rho_{\rm thr}} / \bar \rho_{\rm{b}}^2$ as a function of the
corresponding threshold overdensity $\Delta_{\rm thr}$. This is done by
performing a volume-weighted summation over all SPH particles with
overdensities $\Delta_i < \Delta_{\rm thr}$, i.e. we set $C(<\Delta_{\rm
thr})= {\sum_{\Delta_i < \Delta_{\rm thr}} h_i^3 \Delta_i^2} / {
\sum_{\Delta_i < \Delta_{\rm thr}} h_i^3}$, where $h_i$ is the radius of the
SPH smoothing kernel associated with SPH particle $i$.
\par
By definition, $C(< \Delta_{\rm thr})$ increases monotonically with the
threshold overdensity $\Delta_{\rm thr}$. We set an upper limit $\Delta^*$ to
$\Delta_{\rm thr}$, corresponding to the threshold density $n^*_{\rm H} \equiv
10^{-1}~\mbox{cm}^{-3}$ for the onset of star formation that we employ in our
simulations. This choice is conservative, since the threshold density marking
the escape of ionising photons and hence the clumping factor of the IGM to be
used in Eq.~\ref{Eq:CriticalSfr} is likely lower. We conservatively identify
the clumping factor $C(<\Delta^*)$ with the clumping factor of the IGM, i.e.\
$C_{\rm IGM} \equiv C(<\Delta^*)$.
\par
\begin{figure}
\begin{center}
  \includegraphics[width=0.49\textwidth]{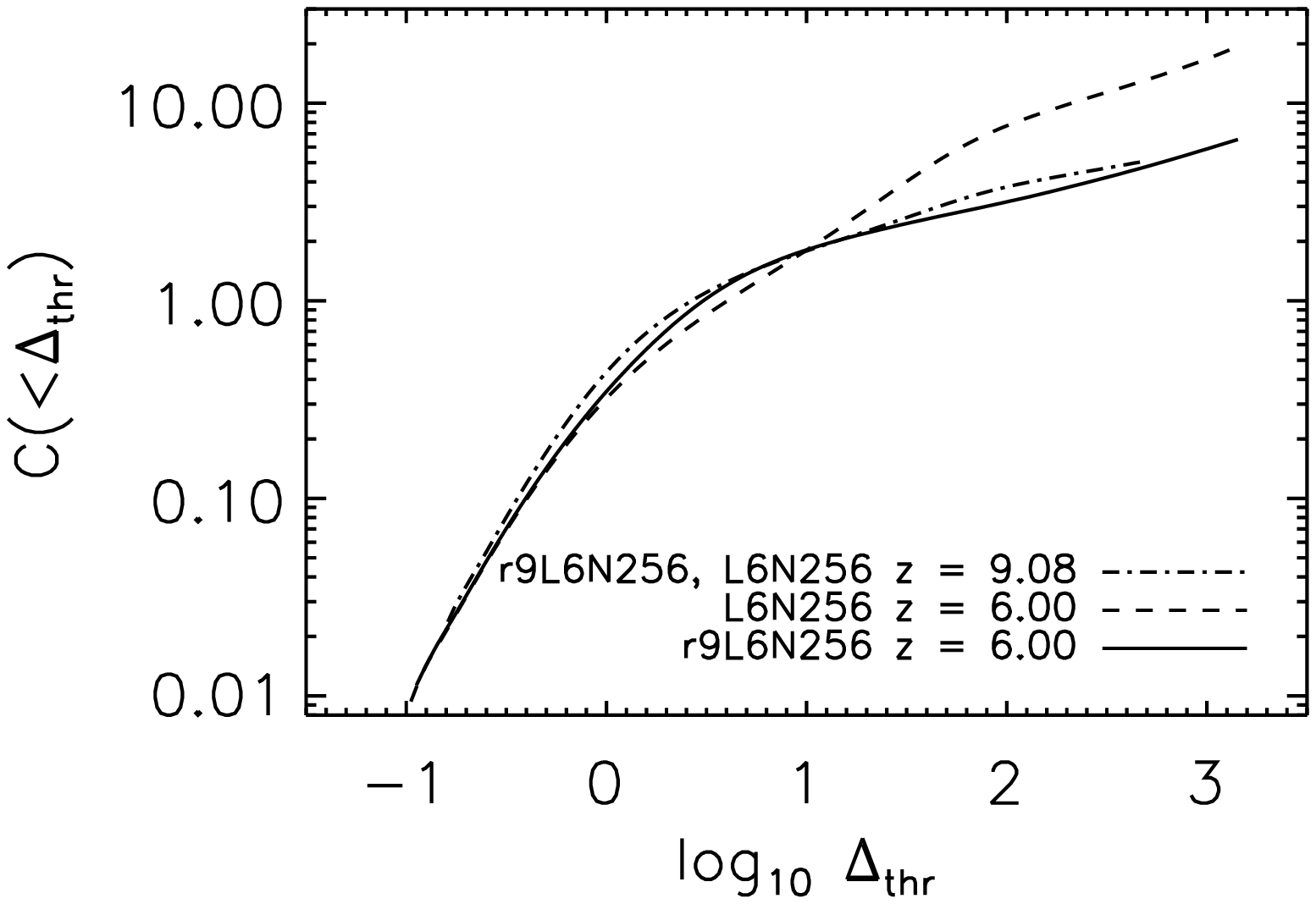}
  \includegraphics[width=0.49\textwidth]{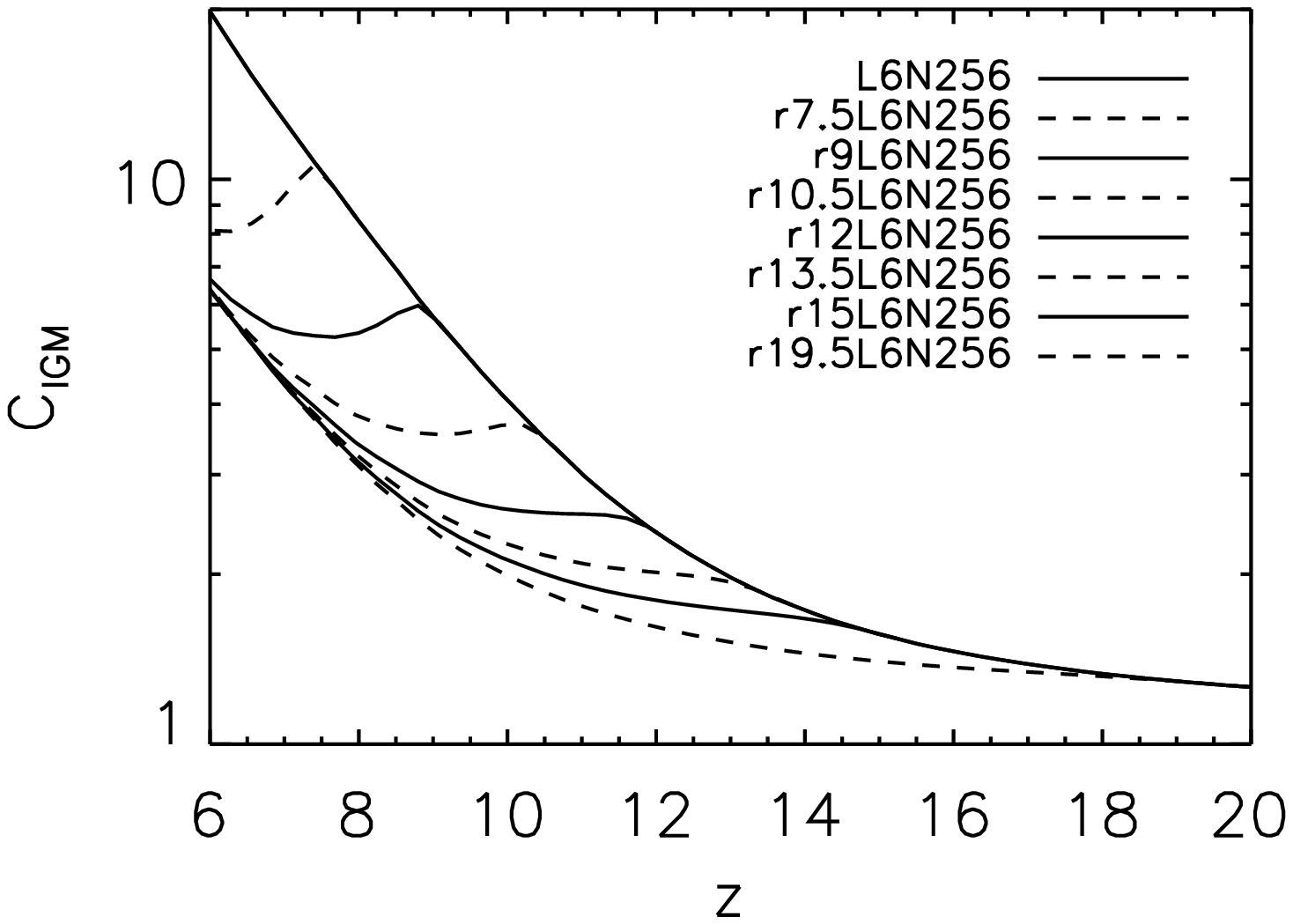}

  \caption{{\it Left-hand panel:} Clumping factor $C(< \Delta_{\rm thr})$ of gas with overdensities 
    $\Delta < \Delta_{\rm thr}$. The inclusion of photoheating
    in \textit{r9L6N256} leads to a clumping factor that is substantially smaller than that obtained from \textit{L6N256},
    for threshold overdensities $\Delta_{\rm thr} > 10$. The maximum threshold
    overdensity we consider corresponds to the critical 
    density $n_{\rm H}^* \equiv 10^{-1} ~\mbox{cm}^{-3}$ (translating into an
    overdensity $\Delta^*$) for the onset of star-formation.
    {\it Right-hand panel:} Evolution of the clumping factor $C_{IGM} \equiv C(< \Delta^*)$ for different 
    reheating redshifts $z_{\rm{r}}$. At $z = 6$ we find $C_{IGM} \approx 6$,
    insensitive to the reheating redshift provided that $z_{\rm r} \ga
    9$. Based on \cite{pawlik_Pawlik:2009a}.}
  \label{Fig:Cmax}
\end{center}
\end{figure}

The left-hand panel of Fig.~\ref{Fig:Cmax} shows $C(< \Delta_{\rm thr})$ for
the simulations \textit{r9L6N256} and \textit{L6N256} at redshifts shortly
before (at $z = 9.08$) and well after (at $z = 6$) the reheating redshift
$z_{\rm{r}} = 9$.  The inclusion of photoheating strongly reduces the clumping
factor for large threshold overdensities. This is because the associated
increase in pressure smoothes small-scale density fluctuations. Note that
because of the reduction of the clumping factor, photoheating provides a
positive feedback on reionisation, which acts in addition to the well-known
negative feedbacks (for a review see, e.g., \citealp{pawlik_Ciardi:2005}). At
$z = 6$ the clumping factor reaches a maximum value of $C(<\Delta^*) \approx
6$. This is five times smaller than the value quoted in
\cite{pawlik_Gnedin:1997}, which has been employed in many (observational)
studies.  The difference with our result is probably because
\cite{pawlik_Gnedin:1997} computed the clumping factor using a density
threshold implicitly set by the maximum overdensity resolved in their
simulation. The right-hand panel of Fig.~\ref{Fig:Cmax} shows the evolution of
the clumping factor $C_{\rm IGM} \equiv C(<\Delta^*)$ for different reheating
redshifts. It demonstrates that our estimate of the clumping factor of the IGM
at $z = 6$, $C_{\rm IGM} \approx 6$, is insensitive to variations in the
redshift at which the UV background is turned on.
\section{Discussion}
\label{Sec:Clumping:Conclusions}
We computed the clumping factor $C_{\rm IGM} \equiv \langle \rho_{\rm{b}}^2
\rangle_{\rm IGM} / \bar\rho_{\rm{b}}^2$, a measure for the average
recombination rate in the intergalactic medium (IGM), using cosmological
smoothed particle hydrodynamics simulations. The clumping factor of the IGM
depends critically on the definition of which gas is considered to be part of
the IGM. Following \cite{pawlik_Miralda:2000}, we assumed that all gas with
densities below a threshold density constitutes the IGM and computed the
clumping factor as a function of this threshold density.
\par
Even our most conservative estimate for the clumping factor, $C_{\rm IGM}
\approx 6$, is five times smaller than the clumping factor that is usually
employed to determine the capacity of star-forming galaxies to keep the $z
\approx 6$ IGM ionised. Setting $C_{\rm IGM} = 6$ in Eq.~\ref{Eq:CriticalSfr},
the critical SFR density is $\dot{\rho}_* = 0.005~f_{\rm esc}^{-1}
~\mbox{M}_{\odot} ~\mbox{yr}^{-1} ~\mbox{Mpc}^{-3}$. This is smaller than
recent observational estimates for the SFR density at $z \approx 6$,
$\dot{\rho}_* \approx 0.02 \pm 0.004 ~\mbox{M}_{\odot} ~\mbox{yr}^{-1}
~\mbox{Mpc}^{-3}$ (integrated to the observed $z \approx 6$ faint-end
limit $L > 0.04 ~L_{z=3}^\ast$ and with new
dust-correction; \citealp{pawlik_Bouwens:2009}), provided that $f_{\rm esc} \ge
0.25$. Our study thus suggests that the observed population of star-forming
galaxies may be capable of keeping the $z \approx 6$ IGM ionised (see also, e.g., \citealp{pawlik_Bolton:2007}). In fact, recent studies suggest that, 
if we assume clumping factors as low as suggested by our work, 
observed galaxies may be capable of keeping the Universe substantially ionised 
out to redshifts as high as $z \la 8$ (e.g., \citealp{pawlik_finkelstein:2009, pawlik_Yan:2009, 
pawlik_Bouwens:2009b, pawlik_Labbe:2009,pawlik_Bunker:2009, pawlik_Kistler:2009}).     
\par
The clumping factor is an important ingredient of (semi-)analytic treatments
of reionisation(e.g.,  \citealp{pawlik_Barkana:2001}).  We have studied the
impact of photoionisation heating on the clumping factor of the IGM assuming a
uniform ionising UV background in the optically thin limit. In reality the
photoheating process will, however, be more complex, with self-shielding being
only one example of the physical effects that we ignored.  The validity of the
approximations inherent to our simplified treatment should be assessed using
high-resolution radiation-hydrodynamical simulations of reionisation that
include cooling by metals and molecules and also allow for feedback from star
formation. For instance, kinetic supernova feedback has been shown to amplify
the effects of photoheating in evaporating the gas out of low-mass halos
(\citealp{pawlik_Pawlik:2009b}), which will affect the clumping factor of the IGM (\citealp{pawlik_Pawlik:2009a}).
\par

\acknowledgements 
This contribution is based on \cite{pawlik_Pawlik:2009a} and is presented with the
kind permission of the publisher Monthly Notices of the Royal Astronomical
Society. The work was supported by Marie Curie Excellence Grant MEXT-CT-2004-014112.

\end{document}